\documentclass[10pt,journal]{IEEEtran}
\usepackage{amsmath}
\usepackage{amsfonts}
\usepackage{graphicx}
\usepackage{float}
\usepackage{tabularx}
\usepackage{amssymb}
\usepackage{pdflscape}  
\usepackage{setspace} 
\usepackage{float} 
\usepackage{color}      
\usepackage{graphicx}        
\usepackage{graphics} 
\usepackage{subfig}                             
\usepackage{epsfig} 
\usepackage{times} 
\usepackage[hyphens]{url}  
\usepackage{comment}
\usepackage{algorithmicx,algorithm}
\usepackage{bbm}
\usepackage{algorithm} 
\usepackage{algpseudocode} 
\usepackage{algorithmicx}
\usepackage{pifont}
\usepackage{cite}
\usepackage{multirow}

\title{A Learning-Driven Stochastic Hybrid System Framework for Detecting Unobservable Contingencies in Power Systems\thanks{The authors are with the Department of Electrical and Computer Engineering, Wayne State University, Detroit, Michigan.
(e-mail: varmazyari.h@wayne.edu; masoud.nazari@wayne.edu). This work is supported in part by National Science Foundation under Grants DMS-2229109.}}
\author{\IEEEauthorblockN{Hamid Varmazyari,~\IEEEmembership{Member,~IEEE,} Masoud H. Nazari,~\IEEEmembership{Senior Member,~IEEE,}}
}
\date{}

\begin{document}
\maketitle
\begin{abstract} 

This paper presents a new learning-based Stochastic Hybrid System (LSHS) framework designed for the detection and classification of contingencies in modern power systems. Unlike conventional monitoring schemes, the proposed approach is capable of identifying unobservable events that remain hidden from standard sensing infrastructures, such as undetected protection system malfunctions. 
The framework operates by analyzing deviations in system outputs and behaviors, which are then categorized into three groups—physical, control, and measurement contingencies—based on their impact on the SHS model. The SHS model integrates both system dynamics and observer-driven state estimation error dynamics. Within this architecture, machine learning classifiers are employed to achieve rapid and accurate categorization of contingencies. The effectiveness of the method is demonstrated through simulations on the IEEE 5-bus and 30-bus systems, where results indicate substantial improvements in both detection speed and accuracy compared with existing approaches. 

\end{abstract}

\section{Introduction}
\label{Sec1}

Modern power systems (MPS) are becoming increasingly complex as they integrate vast numbers of interconnected components, renewable energy resources, advanced communication layers, and distributed control mechanisms. For instance, the Mid-continent Independent System Operator (MISO) grid has
over 16,000 substations, 25,000 buses, and 38,000 lines
\cite{7592475}. 
Traditionally, these systems were designed with an $N-1$ reliability criterion. 
However, in practice, simultaneous or unobservable failures—so-called high-order $N-k$ contingencies—can overwhelm these protections and cause widespread blackouts \cite{climategov, HF3, che2017screening, BISWAS2026112107}. This reality underscores the critical importance of early, accurate contingency detection to maintain system stability and avoid catastrophic disruptions.

A particular challenge arises from unobservable contingencies, which remain undetected under routine measurement and monitoring processes. Such events often manifest only when the system is under stress. 
Examples include failures in protection systems due to misconfigured settings, corrupted communication channels, or unobservable measurement errors \cite{HF1, HF2, zhai2020identifying}. A real-world example is the 2018 Camp Fire in California, initiated by equipment failures in PG\&E’s infrastructure. The incident resulted in tragic human loss and extensive property damage. 

Several classes of unobservable contingencies merit particular attention. Failures in protection systems have been repeatedly identified as key contributors to wide-area disturbances \cite{HF1, ZAKARIYA2023108928, HF5,ghiasi2024enhancing, najafzadeh2024fault}, arising from errors in measurements, misapplied settings, or stress-induced overloads. Similarly, inadequate sensor coverage remains a common limitation in many transmission networks, where line outages or distribution failures cannot be detected in real time \cite{ostrometzky2019physics, sim2024detection}. The increasing interdependence between physical power infrastructures and communication networks introduces additional vulnerabilities to cyber-attacks, false data injection, and other anomalies that can evade detection \cite{liu2016cyber, liu2016false, oshnoei2024smart, shees2024cybersecurity, zhao2024research, lawal2024data}. These diverse sources of unobservable contingencies highlight the need for more comprehensive and robust monitoring solutions.

Over the past decade, multiple methods for contingency detection have been proposed. Statistical techniques \cite{sta1, sta2, sta3, sta4} exploit historical data to identify anomalies but often lack adaptability to new or rare events. Optimization-based approaches \cite{opt1, opt2, opt3} attempt to capture system-wide dynamics but struggle with scalability in real-time applications. Numerical techniques \cite{num1, num2, num3} provide accurate results but are heavily dependent on high-quality data, which is not always available. More recently, artificial intelligence (AI)-based methods \cite{heidari2022accurate, ZAKARIYA2023108928, Guo2023, learn1, 9664189} have shown promise, but their reliance on prior training data and signatures limits their ability to handle contingencies outside the training set. Collectively, these approaches provide valuable insights but still face limitations in detecting unobservable contingencies.

From a mathematical perspective, contingencies can be modeled as discrete events embedded within the continuous dynamics of power system operation\cite{cd1, cd2, yin2022joint, Abadi2025LSHS, Varmazyari2025NAPS}. This duality motivates the use of Stochastic Hybrid Systems (SHS) as a modeling and detection tool. In our earlier work \cite{cd1, cd2}, we developed SHS-based estimation and detection methods for power systems. 
Within this framework, each contingency corresponds to a distinct switching scenario. However, as the number of potential contingencies grows, the number of contingency scenarios increases. For example, 
the MISO’s estimator solves the contingency analysis
application every 5 minutes, and at each run time, it solves more than 10,000 contingencies.
%
Building on this foundation, this paper introduces a learning-based SHS (LSHS) framework to 
integrate machine learning methods into the hybrid modeling framework.
The LSHS incorporates both system dynamics and observer-based state estimation error dynamics. Contingencies are categorized into three domains—physical, control network, and measurement—enabling more structured detection and classification.

The main contributions of this paper are summarized as follows. 1) An LSHS framework is developed for the early detection of contingencies using limited sensing and monitoring data. Contingencies are modeled as discrete events, with changes in the system transfer function serving as indicators. 2)
To improve scalability, a structured classification mechanism is introduced that partitions contingencies into physical, control, and measurement categories, according to their impact on system dynamics. This categorization narrows the search space, reduces computational burden in large-scale systems, and enhances both detection accuracy and detection time. 3) The LSHS formulation is extended to integrate system dynamics with observer state estimation error dynamics, providing robust detection capability across physical, control, and monitoring domains.

The remainder of the paper is organized as follows. Section \ref{Sec2} presents the foundations of SHS modeling and introduces the various types of cyber-physical contingencies along with their impacts on the SHS model. Section~\ref{Sec4} presents the LSHS approach in detail, describing both the modeling enhancements and the machine learning components used for detection and classification. Section~\ref{Sec5} evaluates the effectiveness of the method through comprehensive simulations on the IEEE 5-bus and 30-bus test systems. 
Finally, Section~\ref{Sec6} concludes the paper with a summary of key findings and a discussion of future research directions.


\section{MPS Dynamics Modeling and Contingency Detection in the SHS Framework}
\label{Sec2}

Contingencies introduce sudden changes in the topology and/or parameters of a power system. 
This can be modeled as a stochastic discrete event that causes abrupt jumps in the continuous states of the power system—for example, a sudden frequency drop. The interaction between the continuous dynamics and discrete contingency events gives rise to an SHS model \cite{yin2022joint}. This forms a randomly switching system, which can be modeled as \cite{cd2}:

\begin{align} \label{GeneralRSLSx}
    \dot{x} &= A(\alpha_k) x + B(\alpha_k) u \\ \label{GeneralRSLSy}
    y &= C(\alpha_k) x
\end{align}
where \eqref{GeneralRSLSx} and \eqref{GeneralRSLSy} describe the dynamics and outputs of SHS, respectively.
$\alpha_k$ is the discrete contingency event, where $\alpha_k \in \mathcal{S}=\{1,2,3,\dots,m\} $. $A(\alpha)$ represents system matrix, $B(\alpha)$ is the control input, and $C(\alpha)$ is the measurement system. This forms a randomly switching system, where the linearized model is represented as a Randomly Switching Linear System (RSLS).

In this paper, the LSHS method is applied to transmission networks with synchronous generators; hence, the system dynamics can be effectively represented using RSLS models. It is important to note that active distribution systems with distributed energy resources (DERs) exhibit nonlinear behavior, and therefore, their dynamics are better characterized by a Randomly Switching Nonlinear System (RSNS) model. In our recent works \cite{Abadi2025LSHS,Varmazyari2025NAPS}, we have begun addressing RSNS-based modeling for contingency detection, which will serve as the focus of our future research endeavors. 

The current model primarily assumes linearized dynamics, which limits its ability to fully  capture nonlinear behaviors. Future research will address this limitation by  incorporating distributed energy resources (DERs) into the hybrid model, enabling  the analysis of nonlinear interactions in modern grids. 


\begin{figure}[t]
    \centering
    \includegraphics[width=0.5\textwidth]{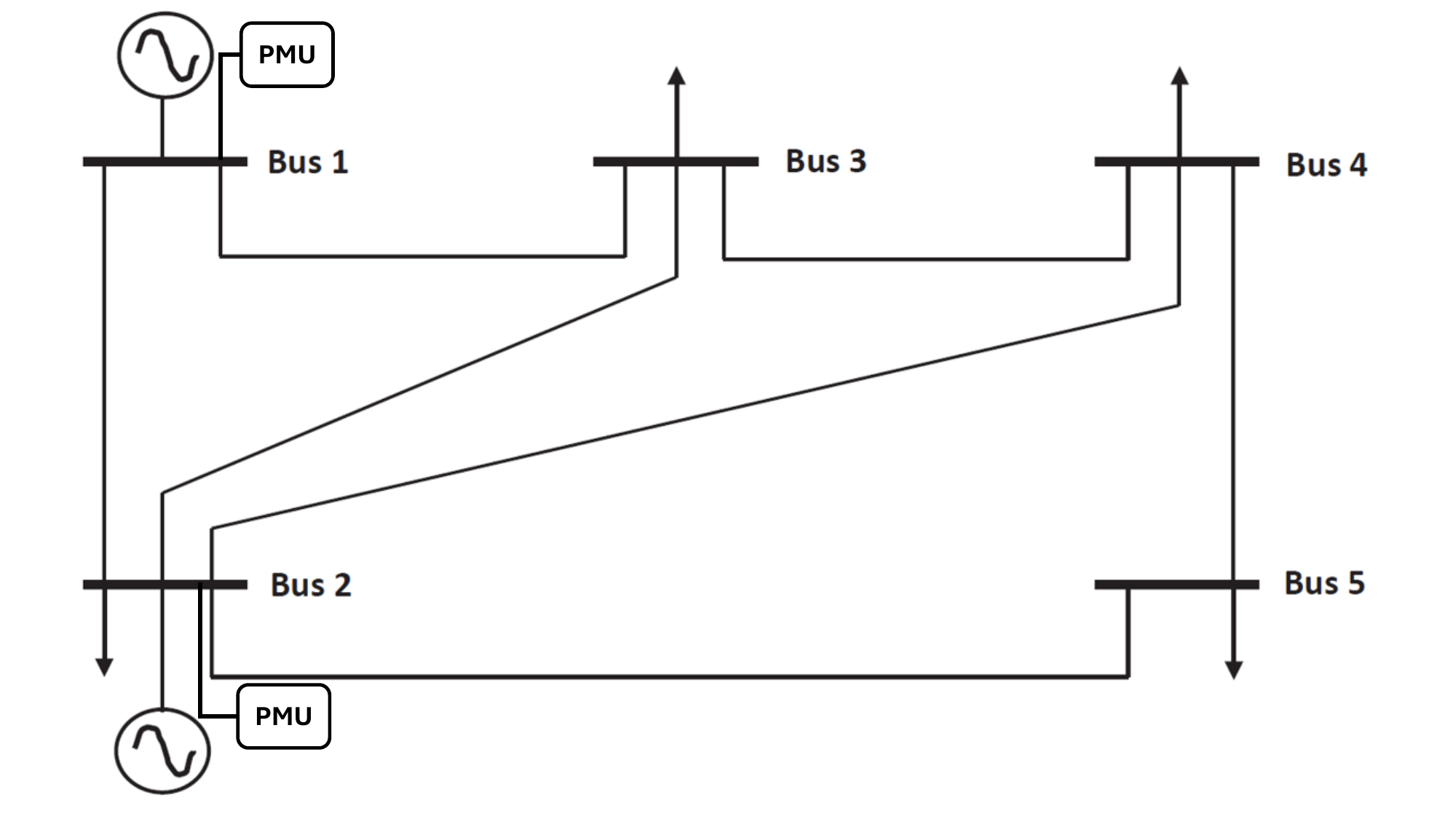}  
    \caption{IEEE 5-bus system.}
    \label{fig:ieee5bus}
\end{figure}





 When a contingency occurs, it typically affects only a specific part of the system, leaving the rest unchanged. As a result, the various switching scenarios of the system that represent these contingencies share common eigenvalues in the matrix 
$A(\alpha)$. 



Contingencies can be categorized into three primary groups based on their impact on the stability and functionality of the power system: physical contingencies 
($\mathcal{S}{p}\subset\mathcal{S}$), control-network contingencies 
($\mathcal{S}{c}\subset\mathcal{S}$), and measurement failures 
($\mathcal{S}_{m}\subset\mathcal{S}$). To illustrate these contingency scenarios, the IEEE 5-bus test system shown in Fig.~\ref{fig:ieee5bus} serves as a representative platform for multi-domain fault analysis.


\subsection{Physical Contingencies}
Physical contingencies are events that directly impact the physical power grid, such as transmission line outages, transformer or generator trips, circuit breaker malfunctions, or failures in protection systems \cite{zhao2019review, sodin2023precise}. Mathematically, these contingencies primarily affect the system matrix 
$A(\alpha)$, expressed as
\begin{equation} {A}(\alpha_k)=\sum_{i=1}^m {A}(i) \mathbbm{1}_{{\alpha_k=i}}, \end{equation}
where 
$\mathbbm{1}_{\alpha}$ is the indicator function of the contingency. In other words, 
$\mathbbm{1}_{\alpha_k} =1$ during the 
$k$th time interval if contingency 
$\alpha_k$ occurs, and 
$\mathbbm{1}_{\alpha_k} =0$ otherwise.

\begin{figure}[h]
    \centering
    \includegraphics[width=1\linewidth]{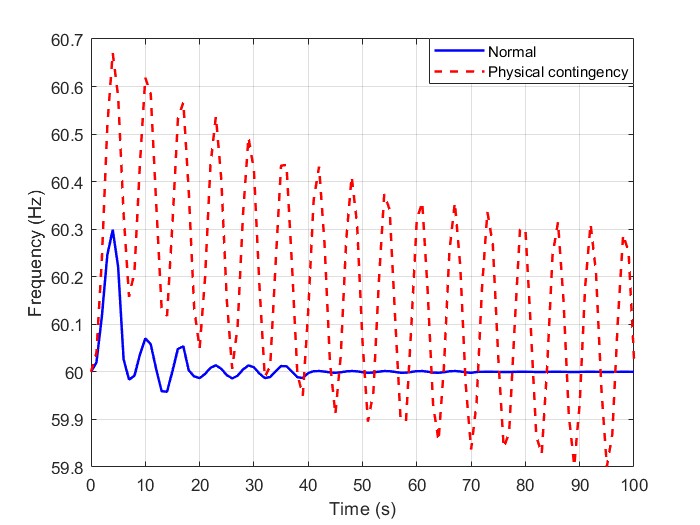}
    \caption{Frequency response of the IEEE 5-bus system at PMU~1 under a Line~1 outage compared with normal operation.}
    \label{fig:phys_cont}
\end{figure}

Physical contingencies events can lead to abrupt changes in network topology or parameters, triggering significant disturbances in power system dynamics. For example, Fig. \ref{fig:phys_cont} shows the dynamic response of the IEEE 5-bus system after the line outage between Buses 1 and 2. 
The line outage changes the network admittance and consequently the system matrix $A(\alpha)$. As observed, during the contingency, the system exhibits an oscillatory behavior with higher amplitude and slower damping compared to the normal condition. This sustained deviation indicates that the topological change in $A(\alpha)$ has shifted the system’s electromechanical modes, reducing damping and imposing greater dynamic stress on frequency stability, even though the overall system remains bounded and stable.

Fast detection of physical contingencies is critical for maintaining power balance, reliability, and cost-effectiveness. 
For example, if during a specific contingency, the system matrix $A(\alpha)$ is not \emph{full-rank}, then equation $A(\alpha) x + B(\alpha) u = 0$ lacks a unique solution, indicating the absence of a feasible power flow solution. This scenario necessitates immediate control actions, such as integrating new energy sources to prevent further deficiencies.

\subsection{Control Network Contingencies}
Control network contingencies encompass unforeseen disruptions affecting the functionality of control systems.
These disruptions stem from communication failures, cyber attacks, human errors, or inaccuracies in signal transmission. The impact of control network contingencies on power systems can be significant and can cause major disruptions. Therefore, rapid and accurate detection of these contingencies is crucial for system stability and resilience. 

This type of contingency can appear in various forms, such as malfunction, denial of service \cite{liu2013denial}, random operation \cite{en14071989}, delay attacks \cite{rouhani2023adaptive}, false data injection \cite{10155170}, packet loss \cite{cetinkaya2016networked}, or oscillatory behavior \cite{FreqOsc}. They interfere with the transmission of accurate information and control signals across the network, leading to erroneous or absent control commands. 

The impact of control network contingencies is modeled in the SHS framework by modifications to the control matrix $B(\alpha_k)$, expressed as: 
\begin{equation} B(\alpha_k)=\sum_{i=1}^m B(i) \mathbbm{1}_{{\alpha_k=i}}. 
\end{equation}

This representation allows for localized failures at node \(i\), impacting \(B_{ii}(\alpha)\) and affecting localized control responses. Similarly, disruptions in communication between nodes \(i\) and \(j\) alter \(B_{ij}(\alpha)\), affecting inter-node control dynamics.

One of the primary challenges is the potential loss of controllability. This risk entails the system’s reduced ability to stabilize or optimize operations during instabilities or other critical situations. To address this, analyzing the controllability matrix is essential:
\begin{equation} \mathcal{C}(\alpha) = \left[B(\alpha) \ \ A(\alpha) B(\alpha) \ \dots \ A(\alpha)^{n-1} B(\alpha)\right]. \end{equation} 
For instance, packet loss in a networked control system disrupts the intended control actions by setting the control input to zero. 
\begin{figure}[t!]
    \centering
    \includegraphics[width=1\linewidth]{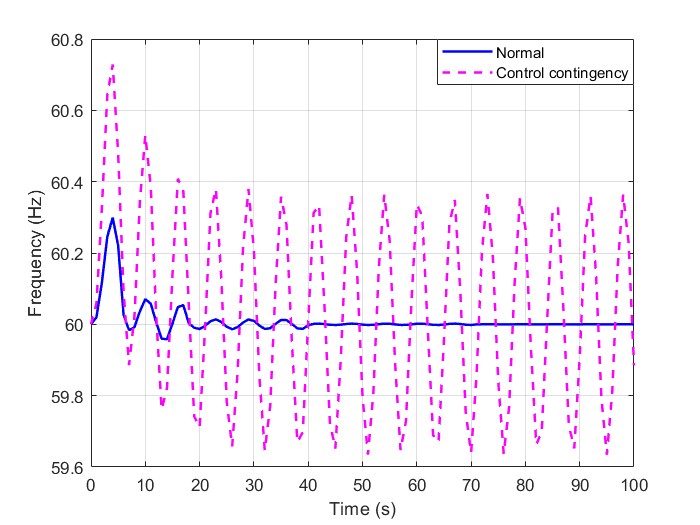}
    \caption{Frequency response of the IEEE 5-bus system at PMU~1 under a control contingency compared with normal operation.}
    \label{fig:ctrl_cont}
\end{figure}

We can mathematically represent this condition by setting the corresponding element in $B(\alpha)$ to zero, directly illustrating that packet loss affects the controllability of the system. As shown in Fig.~\ref{fig:ctrl_cont}, the frequency response of the IEEE 5-bus system measured at PMU~1 is shown for a control contingency scenario in which the control input of Generator~1 is scaled by a factor of~1.20, thereby modifying $B(\alpha)$ and amplifying the control action. This scaling is used to emulate a control-channel failure or misconfiguration,
where the controller applies an incorrect gain to the input signal. 
The contingency scenario exhibits persistent oscillations with greater amplitude and weaker damping compared to the normal condition. 

Although the system remains controllable, control-gain scaling shifts the closed-loop poles closer to the imaginary axis, reducing damping and increasing the frequency deviation from nominal. This highlights how control contingencies can directly compromise system stability.

\subsection{Sensing and Monitoring Contingencies}
Contingencies in the sensing and measurement network represent a critical vulnerability within modern power systems, where inaccurate data is fed into the system's state estimation. These failures can arise from malfunctions or failures of sensors and measurement devices, or from intrusions that corrupt the measurement signals used by the state estimation system. Such discrepancies can severely disrupt operational integrity and lead to catastrophic outcomes. Therefore, the effective detection of measurement contingencies is essential for minimizing their potential impact.

Sensor/monitoring contingencies are modeled in the SHS framework by changes in the matrix $C(\alpha_k)$ as follows:
\begin{equation} C(\alpha_k) = \sum_{i=1}^m C(i) \mathbbm{1}_{{\alpha_k=i}}; 
\end{equation}

\begin{figure}[t!]
    \centering
    \includegraphics[width=1\linewidth]{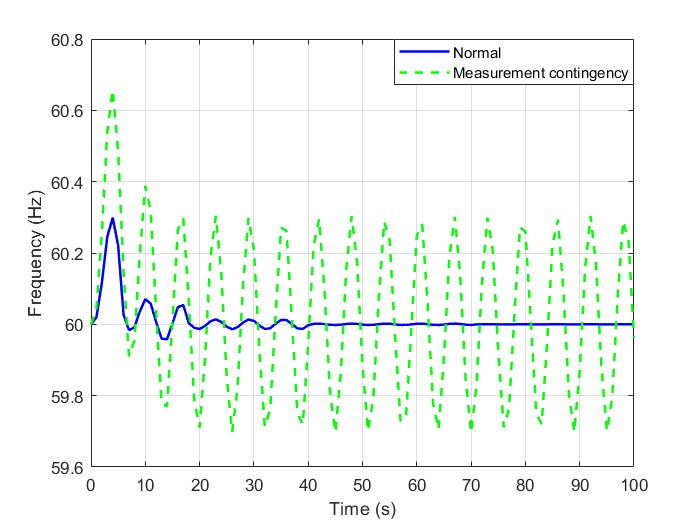}
    \caption{Frequency response of the IEEE 5-bus system at PMU~1 under a measurement contingency compared with normal operation.}
    \label{fig:meas_cont}
\end{figure}

Fig.~\ref{fig:meas_cont} illustrates the effect of a measurement contingency in the IEEE 5-bus system. 
A scaling error of 1.10 is introduced in PMU 1, altering $C(\alpha)$ and distorting the observed frequency output. This scaling emulates a measurement-channel failure,
such as a sensor calibration drift that causes biased readings. The figure shows the frequency measured under the contingency, exhibiting oscillations of higher amplitude and slightly shifted average frequency compared to the normal case. Although the underlying system dynamics remain stable, such deviations can mislead state estimators and obscure the true operating condition. 

This risk is particularly critical because it may not immediately trigger alarms, yet it degrades the accuracy of monitoring and can propagate errors into subsequent control decisions.
Observability is a critical aspect of system design that ensures all system states are accurately estimated. The observability matrix is defined as: \begin{equation} \mathcal{O}(\alpha) = \begin{bmatrix} C(\alpha)^T & (C(\alpha)A(\alpha))^T & \dots & (C(\alpha)A(\alpha)^{n-1})^T \end{bmatrix}^T, \end{equation} which serves as a key criterion for the design of the measurement system and the implementation of backup sensors. The system is typically designed to be observable so that an observer can estimate all system states. However, if a sensor fails, its impact must be reassessed by removing the affected sensor's data from the observability analysis. A single sensor failure can render multiple states of the system unobservable, significantly compromising system monitoring capabilities.
Identifying which portion of the system remains observable under each sensor failure scenario is developed by \cite{cd1}. This understanding shapes the detection algorithm's approach, focusing on the subset of the system that remains observable.

\subsection{Closed-Loop SHS model with Observer Error Dynamics}
 
\begin{figure}[t]
    \includegraphics[width=1.05\linewidth, trim={0cm 0cm 0cm 0cm},clip]{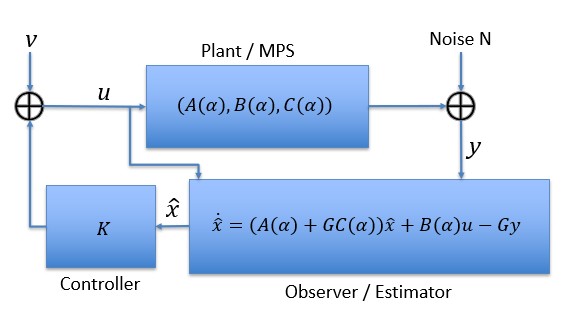}
        \centering
        \captionsetup{width=\linewidth}
        \caption{Closed-loop control scheme with observer-assisted state estimation\cite{lin2007robust}.}
        \label{Fig_ObserverSchematic}
        \vspace{-1.5em}
\end{figure}

 In our earlier work \cite{cd2}, contingency detection was achieved by injecting a probing input into the physical system to identify system anomalies. However, this approach may have limited generalizability, as applying probing inputs is not always feasible for large-scale power systems.
In this paper, we propose an alternative method that integrates observer dynamics and feedback control signals with system outputs within the SHS framework, based on the system described in
 \cite{lin2007robust}. This framework is illustrated in Fig. \ref{Fig_ObserverSchematic}.
For each contingency scenario, this system could be modeled by 
\begin{subequations} \label{eq_ObserverStateSpace}
    \begin{align}
        \label{eq_ObserverStateSpaceA} \dot{x} &= A(\alpha)x + B(\alpha) u \\
        \label{eq_ObserverStateSpaceB} y &= C(\alpha) x + N \\
        \label{eq_ObserverStateSpaceC} \dot{\hat{x}} &= (A(\alpha)+GC(\alpha))\hat{x}+B(\alpha)u-Gy \\
        \label{eq_ObserverStateSpaceD} u &= K \hat{x} + v
    \end{align}
\end{subequations}
where $N$ is the independent and zero-mean Gaussian measurement noise with variance of $\sigma$. Also, $G$ and $K$ are the system state estimation and feedback controller gains. We assume the observer is designed based on a pole-placement approach similar to \cite{lin2007robust}. Accordingly, we can rewrite \eqref{eq_ObserverStateSpaceA} as
\begin{align}
   \notag \dot{x} &= A(\alpha)x + B(\alpha)(K \hat{x} + v) \\
   \notag \quad   &= A(\alpha)x + B(\alpha) K(x -\tilde{x}) + B(\alpha) v \\
   \quad   &= (A(\alpha)+ B(\alpha) K)x - B(\alpha) K\tilde{x} + B(\alpha) v.
\end{align}
Let $\tilde{x}:= x-\hat{x}$ be the estimation error of the observer. Hence, the estimation error dynamics are 
\begin{align}
    \notag \dot{\tilde{x}} &= \dot{x} - \dot{\hat{x}} \\
    \notag \quad   &= A(\alpha)x + B(\alpha) - \left( (A(\alpha)+GC(\alpha))\hat{x}+B(\alpha)u-Gy \right) \\
    \notag \quad   &= \left( A(\alpha)+GC(\alpha) \right) (x - \hat{x}) + GN\\ 
    \quad  &=\left( A(\alpha)+GC(\alpha) \right) \tilde{x} + GN
\end{align}
Thus, we define the closed-loop SHS model as
\begin{align} \label{eq_AdvanceRSLS}
    \begin{bmatrix}
        \dot{x} \\
        \dot{\tilde{x}}
    \end{bmatrix} 
    &=
    \begin{bmatrix}
        A(\alpha) + B(\alpha)K & -B(\alpha)K \\
        \textbf{0} & A(\alpha)+GC(\alpha)
    \end{bmatrix}
    \begin{bmatrix}
        x \\
        \tilde{x}
    \end{bmatrix} \\ \notag
    &+
    \begin{bmatrix}
        B(\alpha) &  \textbf{0} \\ \textbf{0} & G
    \end{bmatrix}
     \begin{bmatrix} v \\ N
    \end{bmatrix},
\end{align}

Since we have access to the state estimation information, we define all $\hat{x}$ values in the closed-loop SHS outputs. Therefore, the output of the system is defined as
\begin{equation} \label{eq_AdvancedSHSoutput}
    y_c = \begin{bmatrix}
        C(\alpha) & \textbf{0} \\ I & -I
    \end{bmatrix} \begin{bmatrix}
        x \\ \tilde{x}
    \end{bmatrix} + \begin{bmatrix}
        N \\ \textbf{0}
    \end{bmatrix}
\end{equation}
where $y_c =[y, \hat{x}]^T$.

\subsection{Modeling Contingencies on Exogenous Signals}
In this section, we examine contingencies that alter the control input (\( u \)) or measurement output (\( y \)) signals of the system. While some contingencies do not directly modify the state-space matrices (\( A \), \( B \), and \( C \)) in the SHS model, they influence the input and output signals. We demonstrate that such contingencies can still be incorporated into the SHS framework and fall into the classes introduced above. 

These types of contingencies which usually stem from cyber-attacks, operational errors, or data handling and processing faults, affect the system's operational dynamics by influencing how parameters are interpreted or utilized. For this type of contingency, we propose considering two equivalent systems based on \eqref{eq_ObserverStateSpace}: the actual system, representing the system under a failure, and the SHS system, which simulates its equivalent behavior as a switching scenario in the SHS model. These systems are defined as follows:
\begin{subequations}\label{H_actual}
\begin{align}
H_{\text{actual}}: & \quad \dot{x} = A_{1}x + B_{1}u, \label{17a}\\
& \quad y = C_{1}x, \label{17b}\\
& \quad \hat{\dot{x}} = (A_{1} + GC_{1})\hat{x} + B_{1}u - Gy.\label{17c},
\end{align}
\end{subequations}

\begin{subequations}\label{H_SHS}
\begin{align}
H_{\text{SHS}}: & \quad \dot{x}' = A_{2}x' + B_{2}u' \\ & \quad y' = C_{2}x' \\ & \quad \hat{\dot{x}}' = (A_{2} + GC_{2})\hat{x}' + B_{2}u' - Gy' ,
\end{align}
\end{subequations}
where the goal is to define the $H_{SHS}$ so that it behaves similar to $H_{actual}$, such that we assume the state space matrices in $H_{\text{actual}}$ (\(A_1, B_1, C_1\)) remain unchanged, reflecting the normal operation of the system. However, parameters such as \(u, y\) change depending on the contingency. The goal for the \(H_{\text{SHS}}\) is to mirror this behavior without altering \(u', y'\) parameters. Instead, we adjust \(B_2, C_2\) accordingly, ensuring that the dynamics represented by \eqref{H_actual} and \eqref{H_SHS} remain consistent across both systems. 

In case of contingencies affecting a control input \(u_i\), the following steps must be carried out:
\begin{enumerate}
    \item Define parameter \(u_i\) under contingency.
    \item Calculate the effect of contingency on \(\dot{x}\), and \(\dot{\hat{x}}\).
    \item Due to the conditions of \(\dot{x}\) = \(\dot{x'}\)  and \(\dot{\hat{x}}\) = \(\dot{\hat{x}}'\), the equality $[B_1]_i u_i = [B_2]_i u'_i$ must hold; where $[B]_i$ represent the $i$th column of $B$.
    \item By defining $[B_2]_i$ = $\frac{[B_1]_i u_i}{u'_i}$, the appropriate $H_{SHS}$ is derived.
 \end{enumerate}
We observe that these unobservable contingencies can be incorporated into the SHS framework as part of the control network contingency by appropriately modifying the matrix $B$.
For example, consider the case of packet loss for one of the control inputs, where \(u_i\) = 0. Thus, \eqref{17a} can be written as $\dot{x} = A_{1}x$. Due to the condition of step 3, we have
\begin{equation}
    A_{1}x = A_{2}x' + B_{2}u'.
\end{equation}
Since this accounts for a control input contingency, we set \(A_{1} = A_{2}\), which leads to \(B_{2} = 0\).
Other types of contingencies may not be as trivial as the packet loss scenario. However, in a normal scenario, we can calculate the effect of the contingencies as a function of the system. 

For a contingency that affects the measurement output \(y_i\), similar steps should be taken, with the following modification: Instead of Step 3 in the control input contingency, \(y'_i = y_i\) must be satisfied, resulting in  
\begin{equation}  
    y_i = [C_{2}]^i x',  
\end{equation}  
where \([C_{2}]^i\) represents the \(i\)th row of the matrix \(C_{2}\). Thus, this contingency falls into the sensing and monitoring class of contingencies.


\section{LSHS Approach for Contingency Detection and Classification}
\label{Sec4}
When the system size is large and the number of contingency scenarios increases, distinguishing and classifying contingencies becomes challenging.
For instance, the method introduced in~\cite{cd2} identifies contingencies by applying probing inputs, observing the system’s output response, and comparing it with real-time measurements over the interval 
\( t \in [k\tau,\,k\tau+\tau_0) \). 
Comparing the system response with the set of all possible $N-1$ and $N-2$ contingency scenarios is time-consuming and becomes computationally intractable for large-scale systems. However, contingency detection must occur within a few seconds to prevent cascading chain reactions and ensure timely corrective actions.

\begin{figure}[t]
    \includegraphics[width=1\linewidth, trim={0cm 0cm 0cm 0cm},clip]{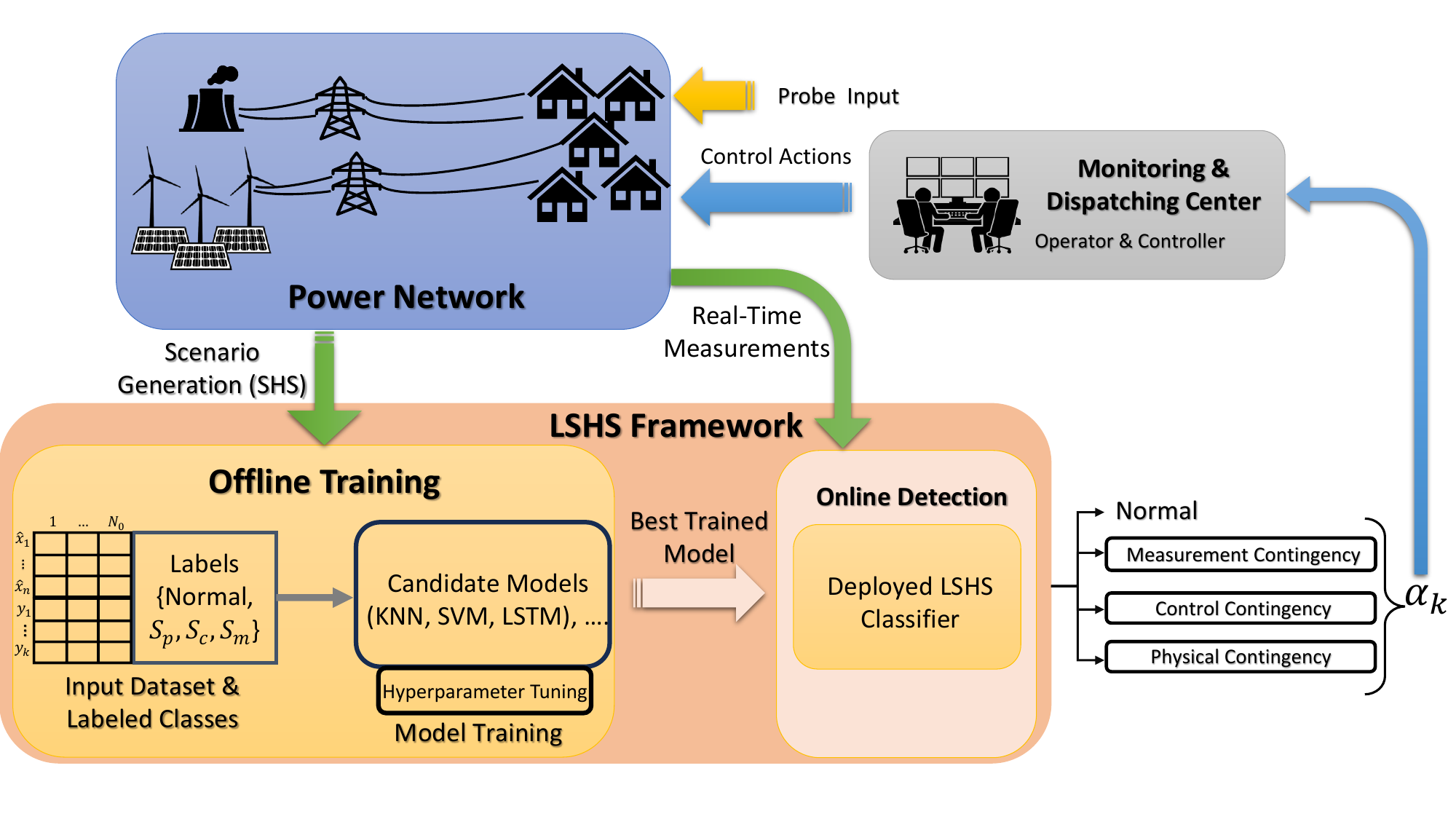}
        \centering
        \captionsetup{width=\linewidth}
        \caption{Proposed LSHS framework for contingency detection and classification. }
        \label{Fig_Classification}
        \vspace{-1em}
\end{figure}

To address this challenge, we propose the application of an ML method for contingency detection and classification, as illustrated in Fig. \ref{Fig_Classification}. This approach leverages the capabilities of ML to effectively reduce the search space, enhance computational efficiency, and improve the accuracy of contingency detection in large-scale systems.

In addition, the ML model can be continuously updated using
historical and newly recorded operational data. By periodically retraining
on this expanding dataset, the model evolves over time to capture the
statistical patterns of new or rare failure events, thereby extending its
classification capability beyond the initially defined contingency categories.

The LSHS method is developed based on the closed-loop SHS model described in \eqref{eq_AdvanceRSLS}. The structure of the system matrix provides critical insights into the contingency classification. 
Let the eigenvalue sets of $(A(i)+B(i)K)$ and $(A(i)+GC(i))$ be $\Lambda_{1,i}$ and $\Lambda_{2,i}$, respectively.
Control contingencies cause variations only in $\Lambda_{1, i}$. 
Sensor/monitoring contingencies cause variations only in $\Lambda_{2, i}$, but Physical contingencies result in changes on both eigenvalue sets.
These structural characteristics are leveraged to label and classify contingency datasets. 

The detection and classification process involves evaluating the error values between \( y_c \) obtained from \eqref{eq_AdvancedSHSoutput} under a contingency, and the nominal system output without any measurement noise effect, denoted as \( y_{\text{nom}} \). 
This error is computed as the element-wise difference between the corresponding outputs:

\begin{equation}
e_i(l,k) = \mid[y_c(\tau k +lt_s)]_i - [y_{\text{nom}}(\tau k +lt_s)]_i\mid,
\end{equation}
where \( e_i(k)=[e(1,k), e(2,k), \dots, e(N_0,k)]^T \) is a vector representing the error values for the $i$th output of the system. 
Assuming 
measurement system $y \in \mathbb{R}^r$ and state estimation $\hat{x}  \in \mathbb{R}^n$, there will be $r+n$ error vectors.
%
The concatenation of error values for each output as $e(k) = [e_1(k), e_2(k), \dots, e_{r+n}(k)]$ could be used as the inputs of the time-series-based classification methods such as long short-term memory (LSTM), where each output is representing one of the features of the classification system. 

For conventional classification methods such as K-nearest neighbors (KNN) or support vector machine (SVM), the sum of error values for each output could be used as the system's features. 
Since the error values could be small, we propose using the logarithm of error values to improve the performance of the classification. 
Additionally, a small value of $\epsilon$ is added to the sum of errors to prevent zero inputs in the logarithm operation. Thus, we have
\begin{equation}
    E_i(k)=log(\sum_{l=1}^{N_0} e_i(l,k) + \epsilon)
\end{equation}
as the aggregation of the error over the $k$th interval.
For classification purposes, the aggregated error \( E_i(k) \) serves as a critical feature to distinguish between different scenarios, and  $E(k) = [E_1(k), E_2(k), \dots, E_{r+n}(k)]$ is used as the input of the classification procedure. 

\begin{figure*}[t]
  \centering
  \includegraphics[width=0.75\textwidth]{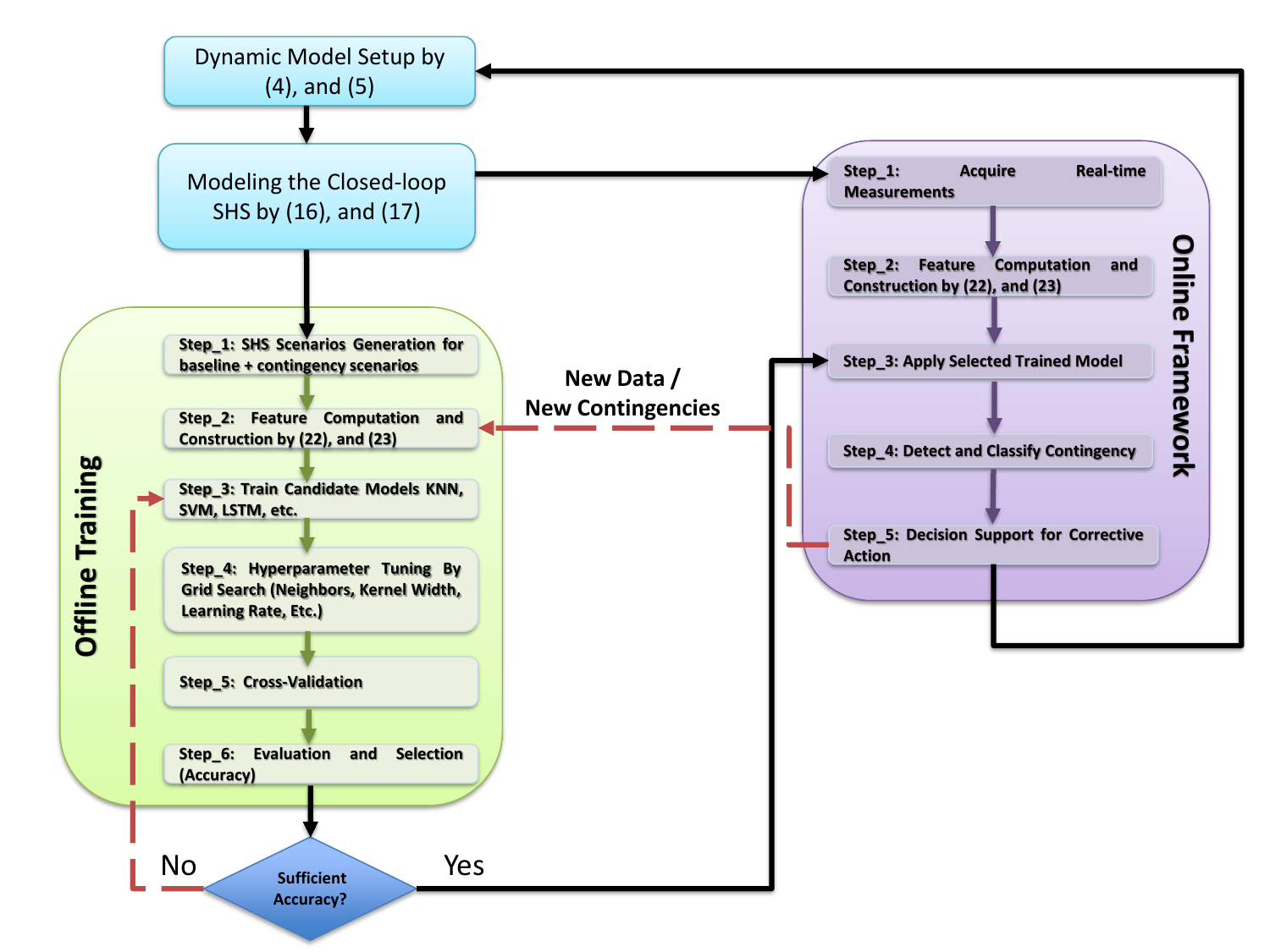}
  \caption{Flowchart of the proposed LSHS-based framework.}
  \label{flowchart}
\end{figure*}

The detection and classification framework, illustrated in Fig.~\ref{flowchart}, is organized into two complementary stages: \textit{offline training} and \textit{online classification}. In the offline training, a comprehensive dataset is generated by simulating a wide range of contingency scenarios. 
For each case, the error values $e_i(l,k)$ are calculated as the difference between the perturbed system output and the nominal reference output. 

The structure of the features depends on the learning method: for LSTM networks, the sequential error values are used directly as time-series input, whereas for KNN and SVM, the aggregated logarithmic error features are employed to ensure numerical stability and improved separability. The candidate learning models are trained on the labeled dataset, and their hyperparameters are optimized through a grid search procedure, where each configuration is evaluated using cross-validation. For KNN this includes varying the number of neighbors and the choice of distance metric; for SVM the penalty parameter $C$, the kernel width $\gamma$, and the classification threshold are optimized; and for LSTM the number of hidden units, layers, dropout rate, learning rate, and batch size are systematically adjusted. The performance of each trained model is compared using accuracy. 

Next, the best-performing model is selected for deployment.
%
Measurement data are continuously collected, and the corresponding error values $e(k)$ and aggregated features $E(k)$ are computed. These features are then processed by the chosen classifier, which directly outputs the predicted contingency type based on the parameters and decision rules determined in the offline stage. The predicted contingency is forwarded to the decision support layer, enabling operators to take timely corrective actions. By confining computationally intensive tasks such as model training, parameter tuning, and threshold calibration to the offline stage, the online stage remains efficient, lightweight, and reliable for real-time operation. This approach enables efficient, accurate, and prompt detection of contingencies as they occur. Importantly, the proposed LSHS framework leverages SHS model outputs to reveal even unobservable contingencies, thereby enhancing detection and classification performance in complex system scenarios.

\section{Simulation}
\label{Sec5}
\begin{figure}[t]
    \includegraphics[width=1\linewidth, trim={0cm 0cm 0cm 0cm},clip]{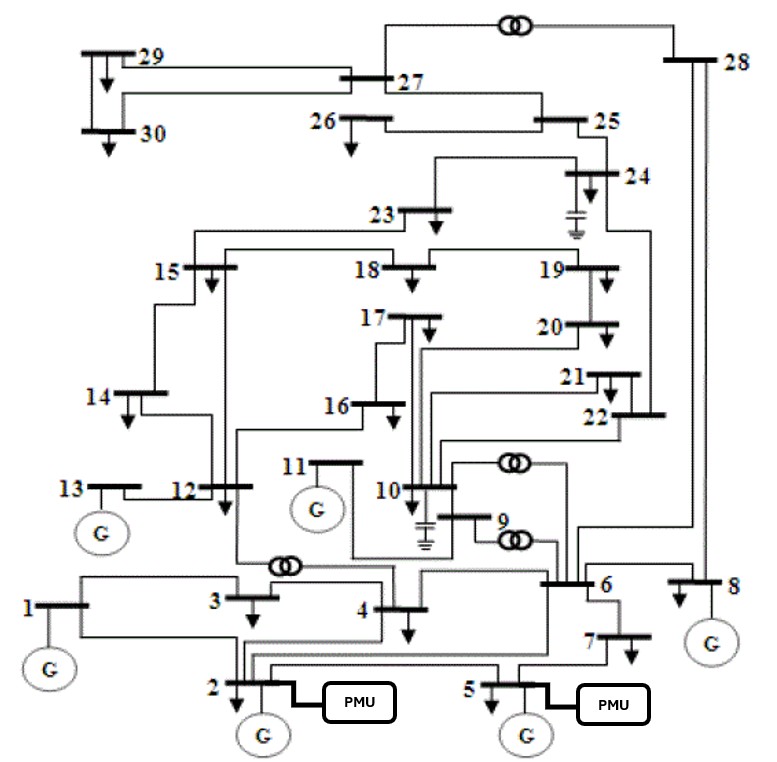}
        \centering
        \captionsetup{width=\linewidth}
        \caption{Single line diagram of standard IEEE-30 bus system\cite{5225036}.}
        \label{IEEE33Bus}
\end{figure}

To evaluate the effectiveness of the proposed LSHS method, we conduct simulations on the modified IEEE 30-bus system, as illustrated in Fig.~\ref{IEEE33Bus}. The detailed system data and line parameters are provided in \cite{5225036}. Bus~1 is modeled as the slack bus with an active power output of approximately $260\,\text{MW}$, while the generator at bus~2 injects about $40\,\text{MW}$. The remaining generators located at buses~5, 8, 11, and 13 operate as PV buses and provide the rest of power.  
The dynamic states of the SHS model are defined for the generators as $x_i = [\delta_i, \omega_i]$ for $i=1,\dots,5$. 
The complete state vector is therefore expressed as $
x = [x_1, x_2, \dots, x_5, \tilde{x}_1, \tilde{x}_2, \dots, \tilde{x}_5]^T.$
PMUs are installed at buses~2 and 5. 

The SHS model has 20 states, with half of the state space representing the physical dynamics, whose eigenvalues are denoted as $\Lambda_1$, and the other half corresponding to the dynamics of the state estimation error, denoted as $\Lambda_2$.
 As described in Section III, physical contingencies influence both $\Lambda_1$ and $\Lambda_2$, while control contingencies primarily affect $\Lambda_1$, and sensor/monitoring contingencies only impact $\Lambda_2$. 
 
 To illustrate the effects of contingencies on system eigenvalues, we present the results in Table~\ref{eigenvalues}. A physical contingency is represented by the line outage between buses 2 and 4, a control contingency is introduced by deviations of the control input $u_3$ by 20\%, and a measurement contingency is modeled as packet loss of $\delta_1$.

\begin{table}[t]
    \centering
    \caption{The closed-loop SHS model eigenvalues under normal operation, physical, control, and measurement contingencies.}
    \label{eigenvalues}
    \resizebox{\columnwidth}{!}{ 
    \begin{tabular}{|c|c|c|c|c|}
        \hline
        & Normal & Physical & Control & Measurement \\ 
        \hline
        \multirow{9}{*}{$\Lambda_1$} & $-2.6768 $ & $-2.5927$ &  $-2.6548$ & $-2.6768$   \\ \cline{2-5}
        & $-2.324 $ & $-2.256$ & $-2.311$ & $-2.324$  \\ \cline{2-5}
        & $-0.7136$ & $-0.6602$ & $-0.7144$ &  $-0.7136$ \\ \cline{2-5}
        & $-0.923 $ & $-0.856$ & $-0.9211$ & $-0.923$  \\ \cline{2-5}        
        & $-1.6778 + j1.2134$ & $-1.6483 + j1.2352$ & $-1.7019 + j1.2093$ & $-1.6778 + j1.2134$ \\ \cline{2-5}
        & $-1.6778 - j1.2134$ & $-1.6483 - j1.2352$ & $-1.7019 - j1.2093$ & $-1.6778 - j1.2134$ \\ \cline{2-5}
        & $-0.3549 + j2.1349$ &$ -0.3733 + j2.0897$ & $-0.3567 + j2.1395$ & $-0.3549 + j2.1349$  \\ \cline{2-5}
        & $-0.3549 - j2.1349$ & $-0.3733 - j2.0897$ & $-0.3567 - j2.1395$ & $-0.3549 - j2.1349$ \\ 
        \cline{2-5}
        & $-0.2322 + j1.3706$ &$ -0.2726 + j1.3893$ & $-0.2141 + j1.3692$ & $-0.2322 + j1.3706$  \\
        \cline{2-5}
        & $-0.2322 - j1.3706$ &$ -0.2726 - j1.3893$ & $-0.2141 - j1.3692$ & $-0.2322 - j1.3706$  \\
        \hline
        \multirow{9}{*}{$\Lambda_2$} & $-15$ & $-26.9432$ & $-15$ & $-30.1281$ \\ \cline{2-5}
        &$ -14$ & $-20.7795$ & $ -14$ &  $-14.0502$ \\ \cline{2-5}
        &$ -13$ & $-12.3286 +j13.3343$ & $ -13$ &  $-12.0706$ \\ \cline{2-5}
        & $-12$ & $-12.3286 -j13.3343$ & $-12$ &  $-10.0865$ \\ \cline{2-5}
        & $-11$ & $-8.6128 + j7.1312$ & $-11$  & $-5.4486$ \\ \cline{2-5}
        & $-10$ & $-8.6128 - j7.1312$ & $-10$ & $ -8.0883$ \\ \cline{2-5}
        & $-9$ & $-3.6321 + j4.0932$ & $-9$ & $-9.1313 + j6.6792$ \\ \cline{2-5} 
        & $-8$ & $-3.6321 - j4.0932$ & $-8$ & $-9.1313 - j6.6792$ \\  
        \cline{2-5}
        & $-7$ &$ -4.6583$ & $-7$ & $-6.0964 + j1.9127$  \\
        \cline{2-5}
        & $-6$ &$ -3.4721$ & $-6$ & $-6.0964 - j1.9127$  \\
        \hline
    \end{tabular}
    } 
\end{table}
For training, the outputs of the SHS are used to construct the dataset, defined as 
$y_c = [\delta_1, \delta_2, \hat{x}_1, \dots, \hat{x}_{10}]$. Samples are generated using the MATLAB 
linear state-space environment based on the SHS model parameters $t_s = 20$~ms and $\tau_0 = 1$~s. 
Unobservable contingencies are identified based on their effects on system dynamics as explained in Section \ref{Sec2}. Such contingencies may be 
simulated using the SHS framework by altering control or sensing channels, or may be obtained from 
historical data. The dataset consists of 960 scenarios, randomly generated with 
240 samples for each class. For physical contingencies, we specifically consider line outage events, 
where one of the transmission lines is disconnected. For control and monitoring contingencies, 
either a control input or a sensor measurement is altered. 
To assess robustness, additive noise is applied at multiple levels to the measured 
outputs, and the difference between the system response under contingency and the nominal output 
without measurement noise is used to form the features. Labels for each data record are assigned 
accordingly. The time-series sequences from each scenario are used to train the LSTM algorithm, while 
the aggregated sequences are used to train the KNN and SVM algorithms. Accuracy is defined as the ratio of correct predictions to the total number of scenario samples.

        \label{Accuracy}
\begin{figure}[t]
    \centering
    \includegraphics[width=1\linewidth]{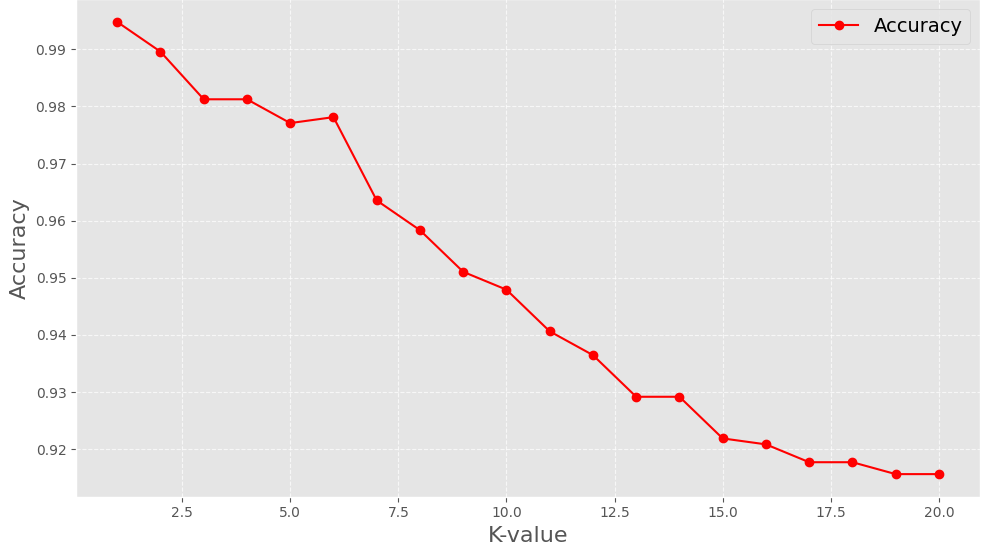} 
    \captionsetup{width=\linewidth}
    \caption{Classification accuracy of the KNN algorithm as a function of the number of neighbors ($k$).}
    \label{KNNtuning}
\end{figure}
\begin{figure}[t]
    \centering
    \includegraphics[width=1\linewidth]{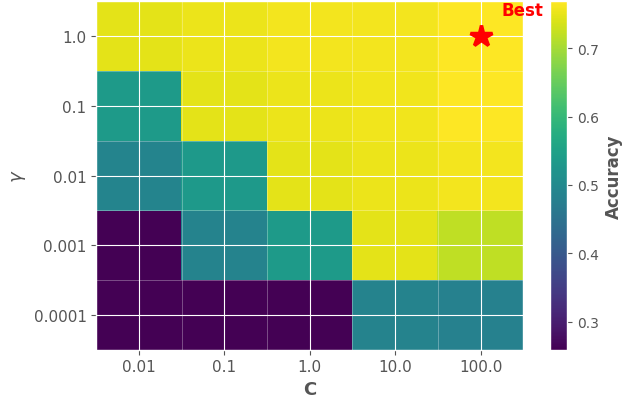} 
    \caption{Grid search results for the SVM classifier with RBF kernel.}
    \label{fig:SVM_heatmap}
\end{figure}

To train each machine learning method, we utilized grid search with 
cross-validation to systematically explore the hyperparameter space and 
select the best-performing configuration. For the KNN classifier, the 
number of neighbors $k$ and different distance metrics were evaluated, 
with the best performance obtained at $k=1$ using the Minkowski distance 
($p=2$), achieving an accuracy of $99.5\%$. The tuning results are 
illustrated in Fig.~\ref{KNNtuning}, which shows the relationship between 
$k$ and classification accuracy.
For the SVM classifier, several kernel 
functions were compared, with the RBF kernel providing the best results. 
A grid search over the penalty parameter $C$ and kernel width $\gamma$ was 
conducted, and the resulting accuracy surface is shown in the heatmap of 
Fig.~\ref{fig:SVM_heatmap}. The best configuration was obtained with $C=100$ and 
$\gamma=1$, with probability calibration and per-class threshold tuning improving the final test accuracy to $90.1\%$. For the LSTM network, 
hyperparameters such as the number of layers, hidden units, dropout rate, 
learning rate, and batch size were tuned, with the best configuration of two 
layers and 128 hidden units yielding a test accuracy of $56\%$. Based on its 
superior performance, the KNN algorithm has been selected 
for implementation in the LSHS framework. 

        \label{KNNconfusion}

To further evaluate its effectiveness, we conduct simulations on the IEEE 30-bus system under a wide range of contingencies. The contingency dataset is organized as follows: $\alpha_k = 1$ to $29$ correspond to $N-1$ and, $N-2$ line outage scenarios; control contingencies are denoted by $\alpha_k = 30$ to $61$, modeled as failures in generator control inputs; and measurement contingencies are defined by $\alpha_k = 62$ to $93$. 
While this study is restricted to 
$N-1$ and $N-2$ contingencies, extending the framework to account for cascading failures caused by higher-order outages, will be considered in future work.

\begin{figure}[t]
    \includegraphics[width=1\linewidth, trim={1.5cm 0cm 2cm 0cm},clip]{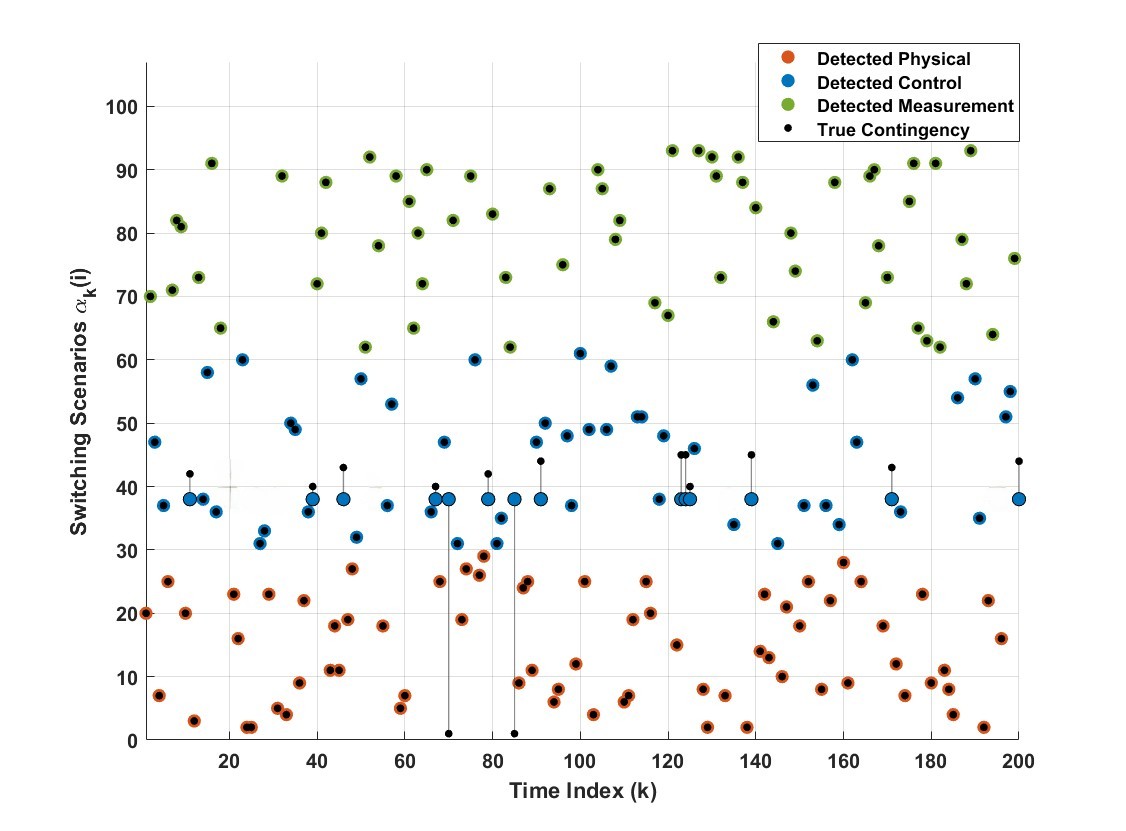}
        \centering
        \captionsetup{width=\linewidth}
        \caption{Random switching of the MPS with contingency detection results obtained using the LSHS method ($\tau_0 = 1$~s).}
        \label{AMPSswitching}
\end{figure}

The detection results obtained using the proposed LSHS framework are shown in Fig.~\ref{AMPSswitching}. It can be observed that the method achieves consistent classification of physical, control, and measurement contingencies. In particular, physical and measurement disturbances are detected with high accuracy, while control-related contingencies remain more challenging because their effects on measurable quantities are indirect and often delayed. Unlike physical faults or sensor anomalies—which immediately alter voltage, current, or frequency measurements—control-related issues primarily influence internal control signals before propagating to the observable system states. Consequently, their transient signatures are more subtle and may be partially masked by normal dynamic variations. Nevertheless, they are still effectively flagged by the LSHS approach.

A quantitative comparison with state-of-the-art methods is provided in Table~\ref{tab:comparison}. As shown, existing approaches such as DQ-GCN (Deep Q-Learning with Graph Convolutional Network)\cite{10436596} and GNN-FNO (Graph Neural Network with Fourier Neural Operator)\cite{11050907} mainly target measurement contingencies, while CBDAC (Conditional Bayesian Deep Auto-Encoder)\cite{9354032} and CNN-CWT (Convolutional Neural Network with Continuous Wavelet Transform)\cite{RIZEAKOS2023120932} are limited to physical failures. The SHS approach~\cite{cd2} is capable of addressing both physical and measurement domains but achieves only 83\% accuracy. By contrast, our proposed LSHS framework uniquely handles all three categories of contingencies with an accuracy of 95-98.33\% at a detection time of $0.02$-$1$~s. This demonstrates both the robustness and scalability of the proposed method, highlighting its superiority over existing approaches in terms of detection speed, accuracy, and coverage of contingency types.


\begin{table}[ht]
\centering
\caption{Comparison of Proposed LSHS with State-of-the-Art Methods}
\label{tab:comparison}
\renewcommand{\arraystretch}{1.2}
\setlength{\tabcolsep}{3pt}
\begin{scriptsize}
\begin{tabular}{|l|p{2.3cm}|c|c|}
\hline
\textbf{Method} & \textbf{Contingency Types} & \textbf{Detection Time (s)} & \textbf{ Accuracy (\%)} \\ \hline
DQ-GCN\cite{10436596} & Measurement & 0.007 & 86.1 \\ \hline
GNN–FNO\cite{11050907} & Measurement & 0.05-0.15 & 92 \\ \hline
CBDAC\cite{9354032} & Physical & -- & 93.43 \\ \hline
CNN-CWT\cite{RIZEAKOS2023120932}& Physical & 0.02-2 & 91-95 \\ \hline
SHS \cite{cd2} & \begin{tabular}[c]{@{}l@{}}Physical, \\ Measurement\end{tabular} &  0.02 & 83 \\ \hline
\begin{tabular}[c]{@{}l@{}}\textbf{Proposed LSHS}\\ \textbf{(this paper)}\end{tabular} 
& \begin{tabular}[c]{@{}l@{}}Physical, Control, \\ Measurement\end{tabular} 
& 0.02-1 & 95-98.33 \\ \hline
\end{tabular}
\end{scriptsize}
\end{table}

\section{Conclusion}
\label{Sec6}

This paper presented a novel LSHS framework for contingency detection and classification in modern power systems. 
By integrating stochastic hybrid system modeling with ML classifiers, the proposed method captures both physical and cyber  behaviors. Contingencies are categorized according to their impact on the power system into three types: physical, control, and measurement contingencies. This structure effectively reduces the search space and facilitates the efficient detection of unobservable contingencies, which cannot be directly captured by the existing sensing infrastructure. 

The effectiveness of the framework was validated through simulations on the 
IEEE 30-bus test system. Using a dataset covering $N-1$, $N-2$ line outages, generator 
control disturbances, and measurement errors, the LSHS algorithm demonstrated 
an  overall accuracy of 95-98.33\% across all contingency classes. 
These results, confirmed by simulations and comparisons with other 
state-of-the-art approaches, highlight the robustness and scalability of the 
proposed method.

%
%
Future research can be focused on advanced 
probabilistic and learning techniques, such as hidden Markov models and deep 
neural networks, to predict cascading chain reactions. These 
extensions will enhance the predictive and resilience-enhancing capabilities of 
the LSHS framework for real-world deployment.

\bibliographystyle{ieeetr}
\bibliography{references}

@article{cd1,
  author       = {Yuan, Shuo and Wang, Le Yi and Yin, George and Nazari, Masoud H.},
  title        = {Stochastic Hybrid System Modeling and State Estimation of Modern Power Systems under Contingency},
  journal      = {arXiv preprint arXiv:2401.16568},
  year         = {2024},
  archivePrefix= {arXiv},
  eprint       = {2401.16568},
  primaryClass = {eess.SY},
  url          = {https://arxiv.org/abs/2401.16568},
  note         = {Accessed: 2025-08-16}
}

@article{cd2,
title = {Contingency detection in modern power systems: A stochastic hybrid system method},
journal = {Sustainable Energy, Grids and Networks},
volume = {39},
pages = {101414},
year = {2024},
issn = {2352-4677},
doi = {https://doi.org/10.1016/j.segan.2024.101414},
url = {https://www.sciencedirect.com/science/article/pii/S2352467724001437},
author = {Shuo Yuan and Le Yi Wang and George Yin and Masoud H. Nazari},
}

@article{yin2022joint,
  title={Joint Estimation of Continuous and Discrete States in Randomly Switched Linear Systems with Unobservable Subsystems},
  author={Yin, George and Lin, Feng and Polis, Michael P and Chen, Wen and others},
  journal={IEEE Transactions on Automatic Control},
  year={2022},
  publisher={IEEE}
}

@article{FreqOsc,
  title={Incorporating demand response with spinning reserve to realize an adaptive frequency restoration plan for system contingencies},
  author={Chang-Chien, Le-Ren and An, Luu Ngoc and Lin, Ta-Wei and Lee, Wei-Jen},
  journal={IEEE Transactions on Smart Grid},
  volume={3},
  number={3},
  pages={1145--1153},
  year={2012},
  publisher={IEEE}
}

@article{zhai2020identifying,
  title={Identifying disruptive contingencies for catastrophic cascading failures in power systems},
  author={Zhai, Chao and Xiao, Gaoxi and Zhang, Hehong and Wang, Peng and Pan, Tso-Chien},
  journal={International Journal of Electrical Power \& Energy Systems},
  volume={123},
  pages={106214},
  year={2020},
  publisher={Elsevier}
}

@article{che2017screening,
  title={Screening Hidden N-$ k $ Line Contingencies in Smart Grids Using a Multi-Stage Model},
  author={Che, Liang and Liu, Xuan and Li, Zuyi},
  journal={IEEE Transactions on Smart Grid},
  volume={10},
  number={2},
  pages={1280--1289},
  year={2017},
  publisher={IEEE}
}

@article{heidari2022accurate,
  title={Accurate, simultaneous and Real-Time screening of N-1, Nk, and N-1-1 contingencies},
  author={Heidari, Hassan and Hagh, Mehrdad Tarafdar and Salehpoor, Pedram},
  journal={International Journal of Electrical Power \& Energy Systems},
  volume={136},
  pages={107592},
  year={2022},
  publisher={Elsevier}
}

@misc{climategov,
  author       = {National Center for Environment Information},
  title        = {Billion-Dollar Weather and Climate Disasters},
  year         = {2024},
  howpublished = {\url{https://www.climate.gov/news-features/blogs/beyond-data/2022-us-billion-dollar-weather-and-climate-disasters-historical}},
  note         = {Accessed: 2024-05-12}
}

@article{HF1,
  title={Review and prospect of hidden failure: protection system and security and stability control system},
  author={Zhao, Lili and Li, Xueming and Ni, Ming and Li, Tianyu and Cheng, Yameng},
  journal={Journal of Modern Power Systems and Clean Energy},
  volume={7},
  number={6},
  pages={1735--1743},
  year={2019},
  publisher={SGEPRI}
}

@article{HF2,
  title={Identification of critical hidden failure line based on state-failure-network},
  author={Li, Linzhi and Liu, Lu and Wu, Hao and Song, Yonghua and Song, Dunwen and Liu, Yi},
  journal={Journal of Modern Power Systems and Clean Energy},
  volume={10},
  number={1},
  pages={40--49},
  year={2021},
  publisher={SGEPRI}
}

@article{HF3,
  title={Screening Hidden N-$ k $ Line Contingencies in Smart Grids Using a Multi-Stage Model},
  author={Che, Liang and Liu, Xuan and Li, Zuyi},
  journal={IEEE Transactions on Smart Grid},
  volume={10},
  number={2},
  pages={1280--1289},
  year={2017},
  publisher={IEEE}
}

@inproceedings{HF5,
  title={Risk assessment of cascading collapse considering the effect of hidden failure},
  author={Salim, Nur Ashida and Othman, Muhammad Murtadha and Musirin, Ismail and Serwan, Mohd Salleh},
  booktitle={2012 IEEE International Conference on Power and Energy (PECon)},
  pages={778--783},
  year={2012},
  organization={IEEE}
}

@article{ZAKARIYA2023108928,
title = {A Systematic Review on Cascading Failures Models in Renewable Power Systems with Dynamics Perspective and Protections Modeling},
journal = {Electric Power Systems Research},
volume = {214},
pages = {108928},
year = {2023},
issn = {0378-7796},
doi = {https://doi.org/10.1016/j.epsr.2022.108928},
url = {https://www.sciencedirect.com/science/article/pii/S0378779622009798},
author = {M.Z. Zakariya and J. Teh},

}

@article{Guo2023, 
author = {Zhenping Guo and Kai Sun and Xiaowen Su and Srdjan Simunovic},
title = {A review on simulation models of cascading failures in power systems},
year = {2023},
journal = {iEnergy},
volume = {2},
number = {4},
pages = {284-296},
url = {https://www.sciopen.com/article/10.23919/IEN.2023.0039},
doi = {10.23919/IEN.2023.0039},
}

@article{zhao2019review,
  title={Review and prospect of hidden failure: protection system and security and stability control system},
  author={Zhao, Lili and Li, Xueming and Ni, Ming and Li, Tianyu and Cheng, Yameng},
  journal={Journal of Modern Power Systems and Clean Energy},
  volume={7},
  number={6},
  pages={1735--1743},
  year={2019},
  publisher={SGEPRI}
}

@article{ostrometzky2019physics,
  title={Physics-informed deep neural network method for limited observability state estimation},
  author={Ostrometzky, Jonatan and Berestizshevsky, Konstantin and Bernstein, Andrey and Zussman, Gil},
  journal={arXiv preprint arXiv:1910.06401},
  year={2019}
}

@article{sim2024detection,
  title={Detection and Location of Open Conductor Faults for Power Distribution Networks Using a Contingency Analysis Approach},
  author={Sim, Gi-Do and Im, Hyo-Seop and Choi, Joon-Ho and Ahn, Seon-Ju and Yun, Sang-Yun},
  journal={IEEE Access},
  year={2024},
  publisher={IEEE}
}

@article{liu2016cyber,
  title={Cyber attacks against the economic operation of power systems: A fast solution},
  author={Liu, Xuan and Li, Zuyi and Shuai, Zhikang and Wen, Yunfeng},
  journal={IEEE Transactions on Smart Grid},
  volume={8},
  number={2},
  pages={1023--1025},
  year={2016},
  publisher={IEEE}
}

@article{liu2016false,
  title={False data attacks against AC state estimation with incomplete network information},
  author={Liu, Xuan and Li, Zuyi},
  journal={IEEE Transactions on smart grid},
  volume={8},
  number={5},
  pages={2239--2248},
  year={2016},
  publisher={IEEE}
}

@article{cetinkaya2016networked,
  title={Networked control under random and malicious packet losses},
  author={Cetinkaya, Ahmet and Ishii, Hideaki and Hayakawa, Tomohisa},
  journal={IEEE Transactions on Automatic Control},
  volume={62},
  number={5},
  pages={2434--2449},
  year={2016},
  publisher={IEEE}
}

@article{sodin2023precise,
  title={Precise PMU-based localization and classification of short-circuit faults in power distribution systems},
  author={Sodin, Denis and Smolnikar, Miha and Rude{\v{z}}, Urban and {\v{C}}ampa, Andrej},
  journal={IEEE Transactions on Power Delivery},
  volume={38},
  number={5},
  pages={3262--3273},
  year={2023},
  publisher={IEEE}
}

@inproceedings{liu2013denial,
  title={Denial-of-Service (dos) attacks on load frequency control in smart grids},
  author={Liu, Shichao and Liu, Xiaoping P and El Saddik, Abdulmotaleb},
  booktitle={2013 ieee pes innovative smart grid technologies conference (isgt)},
  pages={1--6},
  year={2013},
  organization={IEEE}
}

@ARTICLE{10155170,
  author={Yang, Shaohua and Lao, Keng-Weng and Chen, Yulin and Hui, Hongxun},
  journal={IEEE Transactions on Power Systems}, 
  title={Resilient Distributed Control Against False Data Injection Attacks for Demand Response}, 
  year={2024},
  volume={39},
  number={2},
  pages={2837-2853},
  doi={10.1109/TPWRS.2023.3287205}}

@article{rouhani2023adaptive,
  title={Adaptive finite-time tracking control of fractional microgrids against time-delay attacks},
  author={Rouhani, Seyed Hossein and Abbaszadeh, Ebrahim and Sepestanaki, Mohammadreza Askari and Mobayen, Saleh and Su, Chun-Lien and Nemati, Abbas},
  journal={IEEE Transactions on Industry Applications},
  year={2023},
  publisher={IEEE}
}

@Article{en14071989,
AUTHOR = {El-Sheikhi, Farag Ali and Soliman, Hisham M. and Ahshan, Razzaqul and Hossain, Eklas},
TITLE = {Regional Pole Placers of Power Systems under Random Failures/Repair Markov Jumps},
JOURNAL = {Energies},
VOLUME = {14},
YEAR = {2021},
NUMBER = {7},
ARTICLE-NUMBER = {1989},
URL = {https://www.mdpi.com/1996-1073/14/7/1989},
ISSN = {1996-1073},
DOI = {10.3390/en14071989}
}

@book{lin2007robust,
  title={Robust control design: an optimal control approach},
  author={Lin, Feng},
  year={2007},
  publisher={John Wiley \& Sons}
}

@ARTICLE{opt1,
  author={Che, Liang and Liu, Xuan and Li, Zuyi},
  journal={IEEE Transactions on Smart Grid}, 
  title={Screening Hidden N- $k$  Line Contingencies in Smart Grids Using a Multi-Stage Model}, 
  year={2019},
  volume={10},
  number={2},
  pages={1280-1289},
  doi={10.1109/TSG.2017.2762342}}

@ARTICLE{opt2,
  author={Wang, Leibao and Zeng, Siming and Liang, Jifeng and Fan, Hui and Li, Tiecheng and Luo, Peng and Rong, Shiyang and Hu, Bo},
  journal={IEEE Access}, 
  title={Critical Propagation Path Identification for Cascading Overload Failures With Multi-Stage MILP}, 
  year={2022},
  volume={10},
  number={},
  pages={117561-117571},
  doi={10.1109/ACCESS.2022.3218327}}

@ARTICLE{opt3,
  author={Che, Liang and Liu, Xuan and Wen, Yunfeng and Li, Zuyi},
  journal={IEEE Transactions on Reliability}, 
  title={Identification of Cascading Failure Initiated by Hidden Multiple-Branch Contingency}, 
  year={2019},
  volume={68},
  number={1},
  pages={149-160},
  doi={10.1109/TR.2018.2889478}}

@INPROCEEDINGS{num1,
  author={Ma, Xiaohang and Qian, Hongjiang and Wang, Le Yi and Nazari, Masoud H. and Yin, George},
  booktitle={2023 4th Information Communication Technologies Conference (ICTC)}, 
  title={Numerical Solutions for Detecting Contingency in Modern Power Systems}, 
  year={2023},
  volume={},
  number={},
  pages={390-395},
  doi={10.1109/ICTC57116.2023.10154790}}

@ARTICLE{num2,
  author={Li, Linzhi and Liu, Lu and Wu, Hao and Song, Yonghua and Song, Dunwen and Liu, Yi},
  journal={Journal of Modern Power Systems and Clean Energy}, 
  title={Identification of Critical Hidden Failure Line Based on State-failure-network}, 
  year={2022},
  volume={10},
  number={1},
  pages={40-49},
  doi={10.35833/MPCE.2020.000056}}

@ARTICLE{num3,
  author={Jia, Youwei and Xu, Zhao and Lai, Loi Lei and Wong, Kit Po},
  journal={IEEE Transactions on Industrial Informatics}, 
  title={Risk-Based Power System Security Analysis Considering Cascading Outages}, 
  year={2016},
  volume={12},
  number={2},
  pages={872-882},
  doi={10.1109/TII.2015.2499718}}

@ARTICLE{sta1,
  author={Zhou, Kai and Dobson, Ian and Wang, Zhaoyu},
  journal={IEEE Transactions on Power Systems}, 
  title={The Most Frequent N-k Line Outages Occur in Motifs That Can Improve Contingency Selection}, 
  year={2024},
  volume={39},
  number={1},
  pages={1785-1796},
  doi={10.1109/TPWRS.2023.3249825}}

@ARTICLE{sta2,
  author={Vrignat, Pascal and Avila, Manuel and Duculty, Florent and Kratz, Frédéric},
  journal={IEEE Transactions on Reliability}, 
  title={Failure Event Prediction Using Hidden Markov Model Approaches}, 
  year={2015},
  volume={64},
  number={3},
  pages={1038-1048},
  doi={10.1109/TR.2015.2423191}}

@article{sta3,
title = {A robust optimization approach for protecting power systems against cascading blackouts},
journal = {Electric Power Systems Research},
volume = {189},
pages = {106794},
year = {2020},
issn = {0378-7796},
doi = {https://doi.org/10.1016/j.epsr.2020.106794},
author = {Chao Zhai and Hung D. Nguyen and Gaoxi Xiao},
}

@INPROCEEDINGS{sta4,
  author={Albinali, Hussain F and Meliopoulos, A.P.},
  booktitle={2016 IEEE Power and Energy Society General Meeting (PESGM)}, 
  title={A Centralized Substation Protection Scheme that detects hidden failures}, 
  year={2016},
  volume={},
  number={},
  pages={1-5},
  doi={10.1109/PESGM.2016.7741559}}

@ARTICLE{learn1,
  author={Du, Yan and Li, Fangxing and Zheng, Tongxin and Li, Jiang},
  journal={IEEE Transactions on Power Systems}, 
  title={Fast Cascading Outage Screening Based on Deep Convolutional Neural Network and Depth-First Search}, 
  year={2020},
  volume={35},
  number={4},
  pages={2704-2715},
  doi={10.1109/TPWRS.2020.2969956}}

@ARTICLE{7592475,
  author={Li, Xingpeng and Balasubramanian, Pranavamoorthy and Sahraei-Ardakani, Mostafa and Abdi-Khorsand, Mojdeh and Hedman, Kory W. and Podmore, Robin},
  journal={IEEE Transactions on Power Systems}, 
  title={Real-Time Contingency Analysis With Corrective Transmission Switching}, 
  year={2017},
  volume={32},
  number={4},
  pages={2604-2617},
  doi={10.1109/TPWRS.2016.2616903}}

@INBOOK{5225036,
  author={Shahidehpour, Mohammad and Wang, Yaoyu},
  booktitle={Communication and Control in Electric Power Systems: Applications of Parallel and Distributed Processing}, 
  title={Appendix C: IEEE30 Bus System Data}, 
  year={2003},
  volume={},
  number={},
  pages={493-495},
  doi={10.1002/0471462926.app3}}

@ARTICLE{10436596,
  author={Vincent, Edeh and Korki, Mehdi and Seyedmahmoudian, Mehdi and Stojcevski, Alex and Mekhilef, Saad},
  journal={CSEE Journal of Power and Energy Systems}, 
  title={Reinforcement Learning-Empowered Graph Convolutional Network Framework for Data Integrity Attack Detection in Cyber-Physical Systems}, 
  year={2024},
  volume={10},
  number={2},
  pages={797-806},
  doi={10.17775/CSEEJPES.2023.01250}}

@ARTICLE{11050907,
  author={Lu, Genghong and Bu, Siqi},
  journal={IEEE Transactions on Industrial Informatics}, 
  title={Online Power System Dynamic Security Assessment: A GNN–FNO Approach Learning From Multisource Spatial–Temporal Data}, 
  year={2025},
  volume={},
  number={},
  pages={1-11},
  doi={10.1109/TII.2025.3576847}}

@ARTICLE{9354032,
  author={Zhang, Tingqi and Sun, Mingyang and Cremer, Jochen L. and Zhang, Ning and Strbac, Goran and Kang, Chongqing},
  journal={IEEE Transactions on Power Systems}, 
  title={A Confidence-Aware Machine Learning Framework for Dynamic Security Assessment}, 
  year={2021},
  volume={36},
  number={5},
  pages={3907-3920},
  doi={10.1109/TPWRS.2021.3059197}}

@article{RIZEAKOS2023120932,
title = {Deep learning-based application for fault location identification and type classification in active distribution grids},
journal = {Applied Energy},
volume = {338},
pages = {120932},
year = {2023},
issn = {0306-2619},
doi = {https://doi.org/10.1016/j.apenergy.2023.120932},
url = {https://www.sciencedirect.com/science/article/pii/S0306261923002969},
author = {V. Rizeakos and A. Bachoumis and N. Andriopoulos and M. Birbas and A. Birbas}
}

@INPROCEEDINGS{9664189,
  author={Varmaziari, Hamid and Dehghani, Maryam},
  booktitle={2021 11th Smart Grid Conference (SGC)}, 
  title={Cyber Attack Detection in PMU Networks Exploiting the Combination of Machine Learning and State Estimation-Based Methods}, 
  year={2021},
  volume={},
  number={},
  pages={1-6},
  doi={10.1109/SGC54087.2021.9664189}}

@inproceedings{Abadi2025LSHS,
  author    = {E. M. Abadi and H. Varmazyari and M. H. Nazari},
  title     = {Detecting Unobservable Contingencies in Active Distribution Systems Using a Stochastic Hybrid Systems Approach},
  booktitle = {Proc. IEEE Power \& Energy Society General Meeting (PESGM)},
  year      = {2025},
  note      = { available online: \url{https://arxiv.org/abs/2503.02040}},
}

@inproceedings{Varmazyari2025NAPS,
  author    = {H. Varmazyari and M. H. Nazari},
  title     = {A Learning-Based Hybrid System Approach for Detecting Contingencies in Distribution Grids with Inverter-Based Resources},
  booktitle = {Proc. North American Power Symposium (NAPS)},
  year      = {2025},
  address   = {Hartford, CT, USA},
  doi       = {10.48550/arXiv.2508.18500},
  url       = {https://arxiv.org/abs/2508.18500},
  note      = {available online: \url{https://arxiv.org/abs/2508.18500}},
}

@article{BISWAS2026112107,
title = {Data-driven, topology-aware energy resilience assessment and enhancement in urban communities},
journal = {Electric Power Systems Research},
volume = {250},
pages = {112107},
year = {2026},
issn = {0378-7796},
doi = {https://doi.org/10.1016/j.epsr.2025.112107},
url = {https://www.sciencedirect.com/science/article/pii/S0378779625006959},
author = {Antar Kumar Biswas and Hamid Varmazyari and Masoud H. Nazari},
}

@article{ghiasi2024enhancing,
  title={Enhancing power grid stability: design and integration of a fast bus tripping system in protection relays},
  author={Ghiasi, Mohammad and Wang, Zhanle and Mehrandezh, Mehran and Alhelou, Hassan Haes and Ghadimi, Noradin},
  journal={IEEE Transactions on Consumer Electronics},
  year={2024},
  publisher={IEEE}
}

@article{najafzadeh2024fault,
  title={Fault detection, classification and localization along the power grid line using optimized machine learning algorithms},
  author={Najafzadeh, Masoud and Pouladi, Jaber and Daghigh, Ali and Beiza, Jamal and Abedinzade, Taher},
  journal={International Journal of Computational Intelligence Systems},
  volume={17},
  number={1},
  pages={49},
  year={2024},
  publisher={Springer}
}

@article{oshnoei2024smart,
  title={Smart frequency control of cyber-physical power system under false data injection attacks},
  author={Oshnoei, Soroush and Aghamohammadi, Mohammad Reza and Khooban, Mohammad Hassan},
  journal={IEEE Transactions on Circuits and Systems I: Regular Papers},
  volume={71},
  number={12},
  pages={5582--5595},
  year={2024},
  publisher={IEEE}
}

@article{shees2024cybersecurity,
  title={Cybersecurity in smart grids: Detecting false data injection attacks utilizing supervised machine learning techniques},
  author={Shees, Anwer and Tariq, Mohd and Sarwat, Arif I},
  journal={Energies},
  volume={17},
  number={23},
  pages={5870},
  year={2024},
  publisher={MDPI}
}

@article{zhao2024research,
  title={Research on defense strategies for power system frequency stability under false data injection attacks},
  author={Zhao, Zhenghui and Shang, Yingying and Qi, Buyang and Wang, Yang and Sun, Yubo and Zhang, Qian},
  journal={Applied Energy},
  volume={371},
  pages={123711},
  year={2024},
  publisher={Elsevier}
}

@article{lawal2024data,
  title={Data-driven learning-based classification model for mitigating false data injection attacks on dynamic line rating systems},
  author={Lawal, Olatunji Ahmed and Teh, Jiashen and Alharbi, Bader and Lai, Ching-Ming},
  journal={Sustainable Energy, Grids and Networks},
  volume={38},
  pages={101347},
  year={2024},
  publisher={Elsevier}
}

\vspace{-5\baselineskip}
\begin{IEEEbiography}[{\includegraphics[width=1.1in,height=1.6in,clip,keepaspectratio]{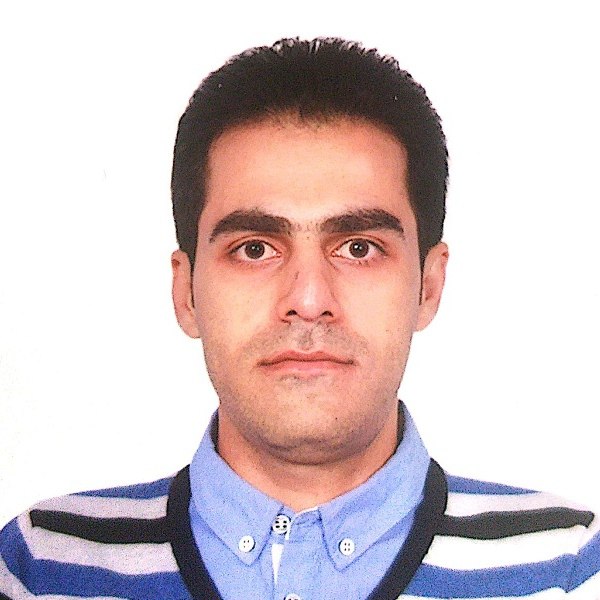}}]{Hamid Varmazyari}
(Member, IEEE) received the B.S. degree in electrical engineering from Bu-Ali Sina University, Hamedan, Iran, and the M.S. degree in electrical engineering from Shiraz University, Shiraz, Iran. He is currently pursuing the Ph.D. degree in electrical and computer engineering at Wayne State University, Detroit, MI, USA. His research focuses on data-driven and stochastic hybrid-system–based methods for resilience assessment and contingency detection in active distribution networks with inverter-based resources. His work has been presented at leading IEEE conferences, including the IEEE Power \& Energy Society General Meeting and the North American Power Symposium (NAPS), and has appeared in peer-reviewed journals such as Electric Power Systems Research. He received the Graduate Student Professional Travel Award from Wayne State University in 2025.
\end{IEEEbiography}
\vspace{-5\baselineskip}
\begin{IEEEbiography}[{\includegraphics[width=1in,height=1.25in,clip,keepaspectratio]{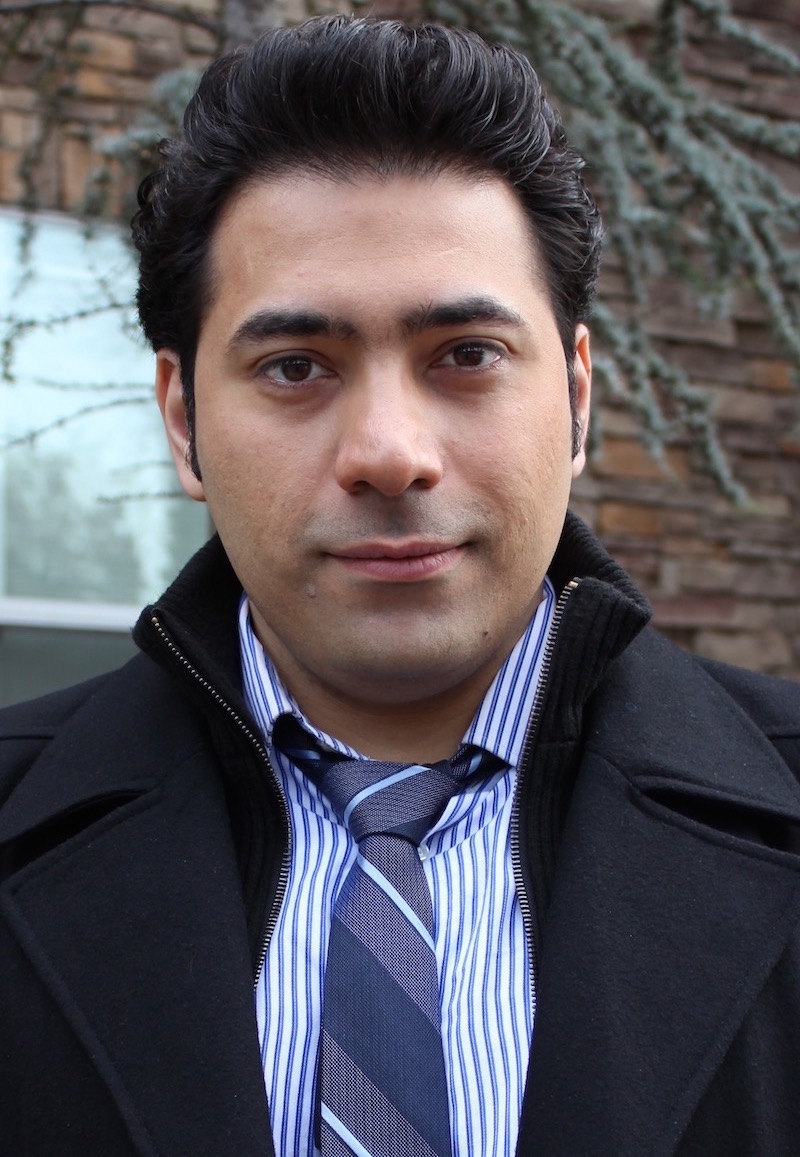}}]{Masoud H. Nazari}
(Senior Member, IEEE) Received the M.S. degree in engineering and public policy from Carnegie Mellon University, the M.S. degree in energy systems from the Sharif University of Technology, and the Ph.D. degree from Carnegie Mellon University.
He was a Postdoctoral Fellow with the School of Electrical and Computer Engineering, Georgia Institute of Technology. Prior to joining Wayne State University (WSU), he was an Assistant Professor of Electrical Engineering with California State University, Long Beach. He is currently an Associate Professor of Electrical and Computer Engineering at WSU and a Global Learning Faculty Fellow with WSU.

He uses new developments in machine learning and distributed control to solve problems in modern power systems (MPS). He has been the Principal Investigator of several research projects supported by the NSF, ARPA-E, DOE, CEC, and DTE Energy. His papers have been published in journals and transactions such as IEEE Transactions on Smart Grid, IEEE Transactions on Power Systems, Automatica, IEEE Transactions on Intelligent Transportation Systems, and IET Generation, Transmission \& Distribution. His research interests include modern power systems, learning, and distributed control.
He received the Best Paper Award from the North American Power Symposium in 2017 and has served as Chair for the IEEE Smart Buildings, Loads, and Consumer Systems Architecture Committee. He also serves on the editorial board for IEEE Technology Conferences.
\end{IEEEbiography}

\end{document}